\newcommand{\Msun}{\ensuremath{{\rm M}_{\odot}}}
\newcommand{\ffrac}[2]{\ensuremath{\frac{\displaystyle #1}{\displaystyle #2}}}

\documentclass{aa}

\usepackage{graphicx}
\usepackage{txfonts}
\usepackage[colorlinks=true,citecolor=blue]{hyperref}
\usepackage{xspace}
\usepackage{mathtools}

\def \s8{{$\sigma_8 \, $}}
\def \om{{$\Omega_{\rm m} \, $}}
\def \logas{{$\log_{10} A_{\rm s} \, $}}

\newcommand{\appropto}{\mathrel{\vcenter{
  \offinterlineskip\halign{\hfil$##$\cr
    \propto\cr\noalign{\kern2pt}\sim\cr\noalign{\kern-2pt}}}}}

\newcommand{\magneticum}{\textit{Magneticum}\xspace}

\newcommand{\agg}[1]{{#1}}
\newcommand{\aggg}[1]{{\bf#1}}

\newcommand{\ed}[1]{{\bf#1}}

\titlerunning{$N_{\textrm s}(M, z, \textrm{cosmology})$}
\authorrunning{Ragagnin et al.}
\begin{document}

   \title{Satellite galaxy abundance dependency on cosmology in Magneticum simulations}


   \author{A. Ragagnin\inst{1,2,3},
           A. Fumagalli\inst{2,3,4},
           T. Castro\inst{2,3,4,5},
           K. Dolag\inst{6,7}, 
           A. Saro\inst{2,3,4,5},
           M. Costanzi\inst{2,3,4},
           S. Bocquet\inst{6}
          }

   \institute{Dipartimento di Fisica e Astronomia "Augusto Righi", Alma Mater Studiorum Università di Bologna, via Gobetti 93/2, I-40129 Bologna, Italy\\
            \email{antonio.ragagnin@unibo.it}
            \and
            INAF - Osservatorio Astronomico di Trieste, via G.B. Tiepolo 11, 34143 Trieste, Italy
            \and
            IFPU - Institute for Fundamental Physics of the Universe, Via Beirut 2, 34014 Trieste, Italy
            \and
            Astronomy Unit, Department of Physics, University of Trieste, via Tiepolo 11, I-34131 Trieste, Italy 
            \and
            INFN - National Institute for Nuclear Physics, Via Valerio 2, I-34127 Trieste, Italy
            \and
            Universitäts-Sternwarte, Fakultät für Physik, Ludwig-Maximilians-Universität München, Scheinerstr.1, 81679 München, Germany 
            \and
            Max-Planck-Institut f\"{u}r  Astrophysik (MPA), Karl-Schwarzschild Strasse 1, 85748 Garching bei M\"{u}nchen, Germany
             }

   \date{submitted}

 
  \abstract
  {{Observational studies as mass-calibrations of galaxy clusters often use mass-richness relations to interpret galaxy number counts.}}
   {To study the impact of {parametrising the richness-mass relation with} cosmological parameters in mock {mass-calibrations, and to understand if this modelling could be inferred by dark-matter only  simulations}.}
   {We build a  Gaussian Process Regression   emulator  of satellite abundance {normalisation and log-slope} based on cosmological parameters $\Omega_m, \Omega_b, \sigma_8, h_0$ and redshift $z.$
   We train our emulator using  \magneticum hydrodynamic simulations that span   different cosmologies \agg{for a given set of feedback scheme paremeters}. }
   {{We find that the normalisation  depends} on cosmological parameters, even if weakly, especially on $\Omega_m,$ $\Omega_b,$ {and that their inclusion in mock-observations increase their constraining power of $10\%.$ }
   {On the other hand the log-slope   is $\approx1$ on every setup, and the emulator  does not predict it with significant accuracy.}  
   We also show that  satellite abundance cosmology dependency differs between full-physics simulations, dark-matter only, and non-radiative simulations.
   }
   { {Mass-calibration studies would benefit by modelling  the mass-richness relations
    with cosmological parameters, especially if this dependency is calibrated from full-physics simulations. }}

   \keywords{Galaxies: clusters: general -- Cosmology: cosmological parameters -- galaxies: abundances -- methods: numerical }

   \maketitle


\section{Introduction}

Properties of galaxies  within   galaxy clusters (GCs)  are connected to the properties of their underlying halo.  This relationship, defined as the Galaxy-Halo (G-H) connection   \citep[see][for a complete review on the topic and its applications]{2018ARA&A..56..435Wechsler}  provides a powerful framework to test galaxy formation models~\citep{2014MNRAS.444..476Reid,2015MNRAS.449.1352Coupon,2017MNRAS.470..651RodriguezPuebla}, to constrain cosmological parameters~\citep{2017MNRAS.467.3024Leauthaud}, and as a proxy to calibrate halo masses \citep{2016MNRAS.462..830Zenteno}.

 One  key topic in the G-H connection is the HOD~\citep[see][for a pioneering study on this topic]{2004ApJ...609...35Kravtsov}, namely the conditional probability distribution $P(N|M)$ that a halo of  mass $M$ has a galaxy abundance $N.$
 In the context of HOD, galaxy counting  is separated into central $N_c$ and  satellite $N_s$ abundances so that:
  \begin{equation}
      N \equiv N_c + N_s.
  \end{equation}

  In fact, central  and satellite galaxies belong to two different populations as they experience different processes~\citep{2002MNRAS.335..311Guzik} as shown by both observations~\citep{2009MNRAS.392.1467Skibba} and numerical simulations~\citep{2018ApJ...864...51Wang}:  the satellite galaxy abundance distribution $P(N_s|M)$ (i.e. the satellite HOD) 
  \agg{is typically modelled with a Poisson distribution} at each mass bin \citep{2004ApJ...609...35Kravtsov} and its average value should increase with halo mass; while the number of  central galaxies  $N_c$ tends to unity asymptotically  with respect to the galaxy mass selection threshold.
       The average $Ns-M$ relation \agg{is typically modelled with a power law at high halo masses}, 
  \begin{equation}
\langle N_\texttt{s}\rangle_M \propto  M ^\beta.
\label{eq:Nsk}
 \end{equation}
 
 \agg{Subhalo population is affected by the host halo accretion history \citep{2008MNRAS.386.2135Giocoli} and HOD normalisation has a mild evolution  with redshift as noted in \cite{2004ApJ...609...35Kravtsov}}.
 The log-slope $\beta$ plays a key role in galaxy formation efficiency and it is not yet well constrained.
 


Constraining the HOD is crucial for interpreting many observational studies \citep[see e.g.][]{2010MNRAS.407..420Ross}, and
   there are efforts to model HOD using additional halo properties besides mass:    assembly bias~\citep{2016MNRAS.460.2552Hearin}, the environment \citep{2020arXiv201204637Voivodic,2021MNRAS.502.3599Hadzhiyska}, a combination of them~\citep{2021MNRAS.502.3582Yuan}, concentration~\citep{2020MNRAS.499.5486Avila},  velocity dispersion~\citep{2021MNRAS.501.1603Hadzhiyska}.
   However, most observational studies deal with a catalogue that have poor or no knowledge about their halo accretion histories{~\citep[see e.g.][]{2019MNRAS.488.4779Costanzi}, part of this work is devoted in understanding if  the dependency of satellite abundance from cosmological parameters can improve mass-calibration studies.
   }


  {There are works in the literature that  study how galaxy populations are affected by variation of cosmological parameters  \citep[see e.g.][]{2005MNRAS.356.1233VanDenBosch,2008MNRAS.384.1301Wang}.
  However, since baryons are known to play a role inside galaxy clusters~\citep{2017MNRAS.469.1997Despali,2021MNRAS.500.2316Castro}, in this work we will show that satellite abundance of  DMO and full-physics (FP) simulations are affected differently from changes in cosmological parameters.}

{We will use} \magneticum\footnote{\href{https://www.magneticum.org}{\url{www.magneticum.org}}} suite of hydrodynamic simulations \citep{2013MNRAS.428.1395Biffi,
2014MNRAS.440.2610Saro, 2015MNRAS.448.1504Steinborn,2015ApJ...812...29Teklu,2015MNRAS.451.4277Dolag,2016MNRAS.463.1797Dolag,2016MNRAS.458.1013Steinborn,2016MNRAS.456.2361Bocquet,2017MNRAS.464.3742Remus,2019MNRAS.486.4001Ragagnin}.
Here we employ a set of $15$ runs  with same initial conditions and run on different cosmological parameters~\citep{2020MNRAS.494.3728Singh,2021MNRAS.500.5056Ragagnin} and were run with the same feedback scheme parameters.

 \agg{The paper is structured as follows: in Sect. \ref{sec:sims}, we describe in detail the numerical set up of the simulations used in this work.
 In Sec. \ref{sec:dmo}, we justify the need of studying HOD cosmology dependency with FP simulations instead of DMO simulations.
 In Sect. \ref{sec:ns}, we fit the satellite abundance for all our simulations and snapshots, build an emulator, \ed{and test it}.
 We devote Sect. \ref{sec:mock} to studying the effect of employing an emulator in mock observations. 
 We draw our conclusions in Sect. \ref{sec:conclu}.}
 

\section{\magneticum Simulations}
\label{sec:sims}

\begin{table*}
\caption{\magneticum simulation specifications used in this work: the Box0/mr, Box1a/mr, Box2/hr, and Box4/uhr. Columns from left to right present,  the name, the cosmology (see Table~\ref{table:2}), the box size in comoving $\mathrm{Mpc}/h,$ dark matter and initial gas particle masses $m_{\mathrm DM}, m_{\mathrm gas},$ and gravitational softening for dark matter, gas and stars $\epsilon_{\mathrm DM}$, $\epsilon_{\mathrm gas}$, $\epsilon_{\mathrm stars}$.}              
\label{table:1}      
\centering                                      
\begin{tabular}{l c c c c c c c}          
\hline\hline   
Name & Cosmologies  & Box size & $m_{DM}$  & $m_{gas}  $ & $\epsilon_{DM}  $ & $\epsilon_{gas}$ & $\epsilon_{stars} $ \\
 &  & $[a\mathrm{Mpc}/h]$ &  $\Bigl[M_\odot/h\ffrac{\Omega_m}{\Omega_{m,WMAP7}}\Bigr]$ & $  \Bigl[M_\odot/h\ffrac{\Omega_b}{\Omega_{b,WMAP7}}\Bigr]$ & $ [a\mathrm{kpc}/h]$ & $ [a\mathrm{kpc}/h]$ & $ [a\mathrm{kpc}/h]$ \\
\hline
Box0/mr & C8 & 2688 & $1.3\cdot10^{10}$ & $2.6\cdot10^9$ & 10 & 10& 5\\
Box1a/mr & C1-15 & 896 & " & " & " & " & " \\
Box2/hr & C8 & 352  & $6.9\cdot10^8$ & $1.4\cdot10^8$ & 3.75 & 3.75 & 2 \\
Box4/uhr& C8 &  48 & $3.6\cdot10^7$ & $7.3\cdot10^6$ & 1.4  & $1.4$ & $0.7$\\
\hline
\end{tabular}
\end{table*}

\begin{table}
\caption{List of \magneticum cosmologies for the Box1a/mr C1--C15 simulations. Starred row (C8$^\star$) represents the original runs on WMAP7 cosmological parameters. Columns from left to right present,  the name, and cosmological parameters $\Omega_m, \Omega_b, \sigma_8, h_0$, respectively. }              
\label{table:2}      
\centering                                      
\begin{tabular}{l c c c c}          
\hline\hline   
Name & $\Omega_m$ & $\Omega_b$ & $\sigma_8$ & $h_0$ \\
\hline
C1 & 0.153 & 0.0408 & 0.614 & 0.666 \\
C2 & 0.189 & 0.0455 & 0.697 & 0.703 \\
C3 & 0.200 & 0.0415 & 0.850 & 0.730 \\
C4 & 0.204 & 0.0437 & 0.739 & 0.689 \\
C5 & 0.222 & 0.0421 & 0.793 & 0.676 \\
C6 & 0.232 & 0.0413 & 0.687 & 0.670 \\
C7 & 0.268 & 0.0449 & 0.721 & 0.699 \\
C8$^\star$ & 0.272 & 0.0456 & 0.809 & 0.704\\
C9 & 0.301 & 0.0460 & 0.824 & 0.707 \\
C10 & 0.304 & 0.0504 & 0.886 & 0.740\\
C11 & 0.342 & 0.0462 & 0.834 & 0.708\\
C12 & 0.363 & 0.0490 & 0.884 & 0.729\\
C13 & 0.400 & 0.0485 & 0.650 & 0.675\\
C14 & 0.406 & 0.0466 & 0.867 & 0.712\\
C15 & 0.428 & 0.0492 & 0.830 & 0.732\\
\hline
\end{tabular}
\end{table}

\magneticum simulations are based on the  N-body code \texttt{P-Gadget3}, which is an improved version of \texttt{P-Gadget2} \citep{2005Natur.435..629Springel,2005MNRAS.364.1105Springel,2009MNRAS.398.1150Boylan}, with a space-filling curve aware neighbour search \citep{2016pcre.conf..411Ragagnin}, and an improved Smoothed Particle Hydrodynamics (SPH)  solver \citep{2016MNRAS.455.2110Beck}.
These simulations include a treatment of radiative cooling, heating, ultraviolet (UV) background,  star formation and stellar feedback processes as in \cite{2005MNRAS.361..776Springel} connected to a detailed chemical evolution and enrichment model as in \cite{2007MNRAS.382.1050Tornatore}, which follows  11 chemical  elements (H, He,
C, N, O, Ne, Mg, Si, S, Ca, Fe) with the aid of CLOUDY photo-ionisation
code \citep{1998PASP..110..761Ferland}.  \cite{2010MNRAS.401.1670Fabjan,2014MNRAS.442.2304Hirschmann} describe prescriptions for black hole growth and for feedback from AGNs.

Haloes together with their member galaxies are identified using respectively, the FoF halo finder~\citep{1985ApJ...292..371Davis} and  an improved version of the subhalo finder  SUBFIND \citep{2001MNRAS.328..726Springel},   that takes into account the presence of baryons \citep{2009MNRAS.399..497Dolag}.

In this work,  we mainly focus on  a set of $15$ simulations labelled Box1a/mr C1--C15 simulations. They span a range of total matter fraction $0.153<\Omega_m<0.428,$ baryon fraction $0.0408<\Omega_b<0.0504,$ power spectrum normalisation $0.650<\sigma_8<0.886,$ and reduced Hubble constant $ 0.670<h_0<0.732, $
as presented in Tables~\ref{table:1} and~\ref{table:2}, and are   centered around the one of C8, that has WMAP7 cosmological parameters.
For each simulation we study the haloes at a time slice with redshifts $z=0.00,\ 0.14,\ 0.29,\ 0.47.$ 

In order to study resolution and mass-range of our emulator,  we will use three additional \magneticum simulations, all with the same WMAP7 cosmology as C8: we use a high-resolution (HR) simulation Box2/hr\footnote{Box2/hr haloes data is available in the web portal presented in \cite{2017A&C....20...52Ragagnin}}~\citep{2014MNRAS.442.2304Hirschmann}; 
we use a ultra-high resolution  simulation Box4/uhr~\citep{2015ApJ...812...29Teklu} to study the emulator mass range validity on low-mass haloes; and  a large-volume MR simulation \citep[Box0/mr,][]{2016MNRAS.456.2361Bocquet} in order to validate our satellite HOD results up to the most massive galaxy clusters
of the Universe. 
{Note that the phases of the initial conditions of these three boxes are different. }

In this works, all masses and radius are expressed in physical units (unless in Table~\ref{table:1} where the units has been chosen differently for the sake of conciseness), thus they are not implicitly divided by $(1+z)$ or $h_0$ as other works on simulations.

\section{DMO vs. FP simulations}
\label{sec:dmo}

{In this section we demonstrate the need of employing FP simulations (as opposed to DMO ones)   in studies that aim to model the mass-richness relation as a function of cosmology.  In particular we will show  that FP $N_s$ cosmology dependency differs from adiabatic runs (i.e. runs with no cooling and star formation, hereafter {\it norad}) or DMO simulations.}

\begin{figure}
   \centering
   \includegraphics[width=0.5\textwidth]{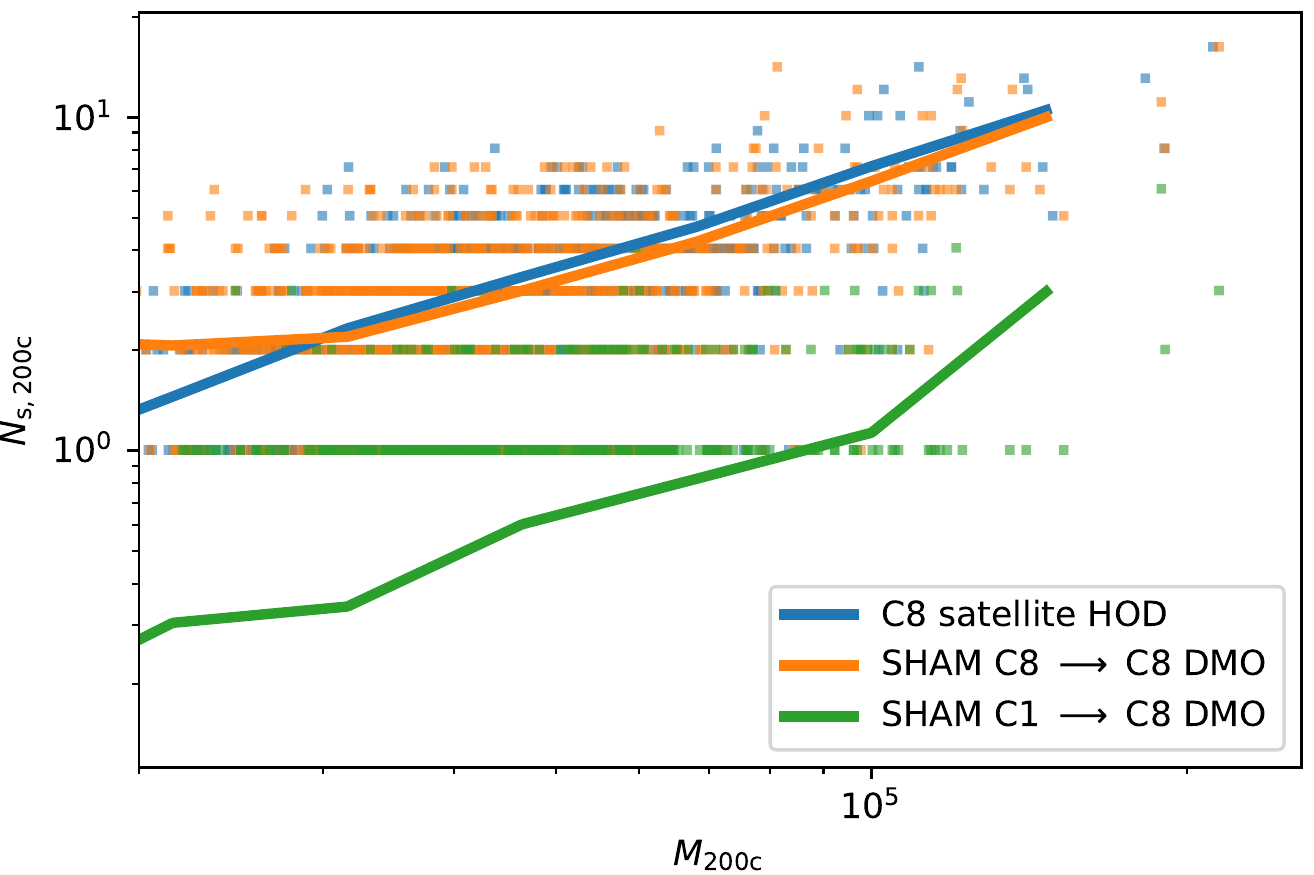}
   \caption{Satellite count $N_{\texttt s, 200c}$  $vs.$ halo mass $M_{\texttt 200c},$ respectively for C8 (blue coloured data), SHAM C8 DMO (orange coloured data), SHAM C8 DMO from C1 (green coloured data). Each data point represents a halo, solid lines represent the mean abundance per mass bin.  }
    \label{fig:sham}%
\end{figure}

To this end, we first show that satellite HOD cannot be recovered in a DMO simulation using Subhalo Abundance Matching (SHAM)    technique from a FP simulation that was run on a different cosmology.
For this purpose we perform mock SHAMs (with a stellar mass cut $M_\star>2\cdot10^{11}M_\odot$) to C8 DMO both from its FP counterpart C8, and from C1. 
\agg{Figure \ref{fig:sham} shows the resulting $N_s-M$ relations for these setups,  and as expected, we find that SHAM from C8 to C8 DMO does match the original C8 FP average values. Additionally, we see that using C1 to perform  SHAM on C8 DMO leads to a too low $N_s$.}

\begin{figure}
   \centering
   \includegraphics[width=0.49\textwidth]{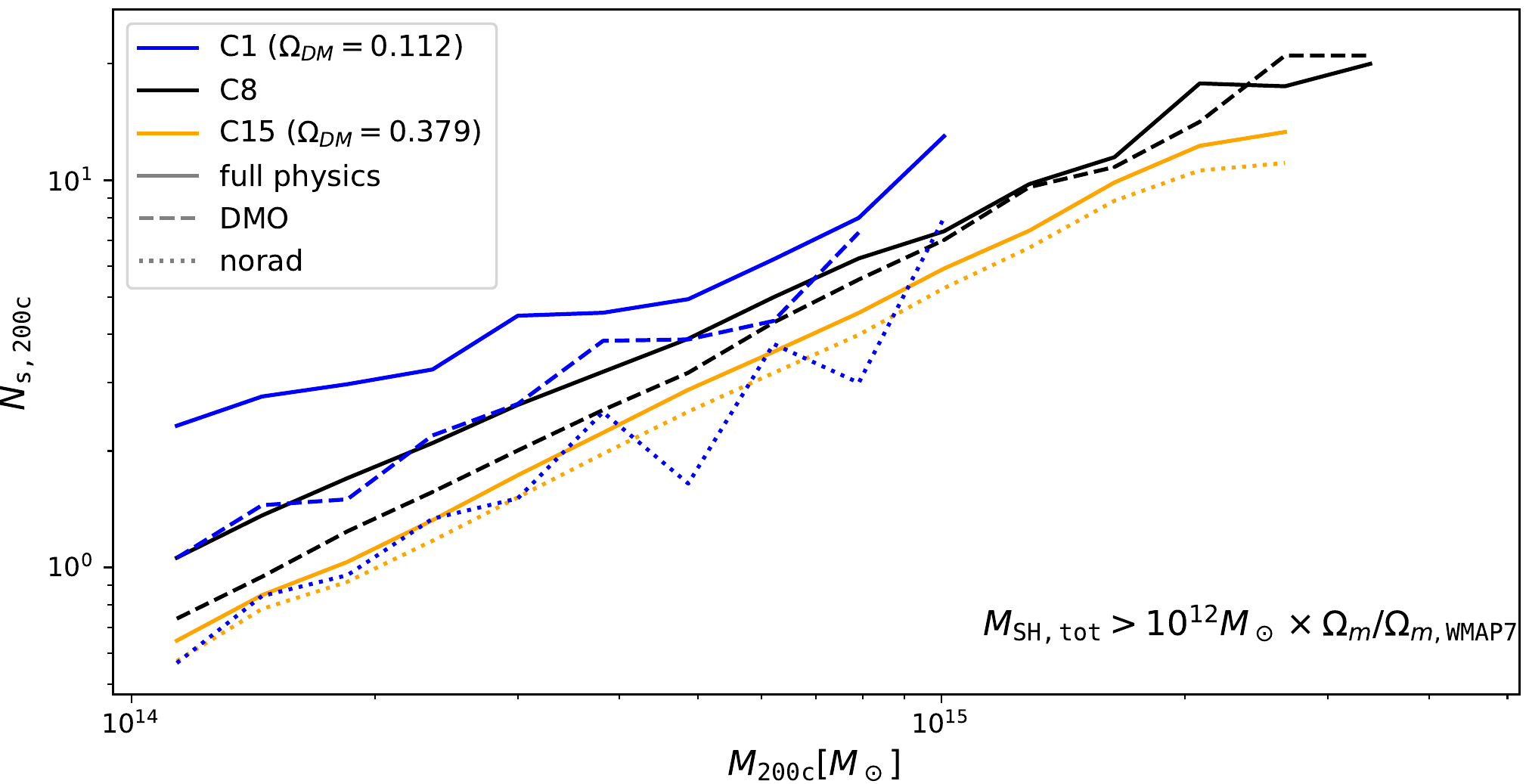}
   \caption{Satellite count $N_{\texttt s, 200c}$  $vs.$ halo mass $M_{\texttt 200c}$ for Box1a/mr C1 (blue higher lower lines), C8 (black intermediate lines) and C15 (orange lower lines) simulations, their non-radiative (dotted lines, abbreviated with {\it norad}) counterpart, and their DMO counterparts (dashed lines). For galaxy masses $M_{\texttt SH,tot}$ greater than $10^{12}\Msun \times \Omega_{m}/ \Omega_{m,{\textrm WMAP7}}.$}
    \label{fig:norad}%
\end{figure}

The reason of this mismatch  lies in the fact that cosmological parameters have a different impact on sub-haloes evolution of  DMO runs and on galaxy evolution of FP runs.
To prove this statement, {in Figure \ref{fig:norad} } we compute $\langle N_s \rangle_M$ for two cosmologies of which we have the {\it norad} counterparts, Box1a/mr C1   and C15, and for two cosmologies of which we have the DMO counterpart, C8 and C1.
Since DMO and {\it norad} simulations have no stars, here we count galaxies based on their total mass $M_{\rm GAL,tot}>10^{12}\Msun,$ re-scaled by $\Omega_m/\Omega_{m, \rm WMAP7}.$

First of all, we note that C8 DMO $Ns-M$ relation is systematically steeper than the respective FP run, thus these kind of relations are strongly affected by the presence of baryons.
Additionally, we can see that  C15 and its   {\it norad} counterpart have almost the same satellite count, while C1 and its  {\it norad} counterpart differ of more than a factor of two, thus the effect of baryons on the $N_s-M$ relation does depend on cosmological parameters.
These two experiments shows that studying HOD dependency on non-FP simulations would have produced a different cosmology dependency than the one found in this work.

\section{Satellite abundance emulator}
\label{sec:ns}

{
In this section we will  train an emulator in order to extrapolate the mass-richness relation for some arbitrary cosmological parameters (that are within the range of the parameters of our simulations). 
We will use this emulator in the next Section, where  we will estimate the benefit of using it in mock mass-calibration studies.
To this end, we first searched for a stellar-mass cut that makes the richness of Box1a/mr C8 to converge with its high-resolution counter-part Box2b/hr (see Appendix \ref{ap:conv} for more details), and found that if we limit ourselves to galaxies with $M_\star>2\cdot10^{11}\Msun,$ then the two simulations have the same mass-richness relations. This relatively-high stellar-mass cut could introduce a bias in the satellite population, however this mass-range should still be enough to constrain the normalisation and log-slope of a power-law modelling.} 

We found that Box2/hr satellite HOD offset between its fiducial stellar mass cut ($10^{10}\Msun$) and the C8 stellar mass cut ($2\cdot10^{11}\Msun$) is 

\begin{equation}
\ffrac{N_{\texttt s}\left(M_\star>10^{10}\Msun\right) }{N_{\texttt s }\left(M_\star>2\cdot10^{11}\Msun\right) } \approx 31.
\label{eq:Nsr}
\end{equation}
We consider this {ratio} useful to compare emulated satellite abundance values (based on C1--C15 MR simulations) predictions together with  HR simulations in literature with a cut $M_\star>10^{10}\Msun$.

To  estimate the satellite count and compare it consistently between different cosmologies, one must choose a minimum stellar mass cut for each set of cosmological parameters.
\agg{ Following  \cite{2020MNRAS.495..686Anbajagane} we decided to re-scale the satellite stellar mass cut with $M_{\star} > 2\cdot10^{11}\Msun \cdot f_b/  f_{b,\texttt{WMAP7}}.$}
 In order to keep the satellite HOD in the power-law  regime, we imposed a halo mass cut so that a given mass bin has at least one halo with $8$ satellites.
 {After this cut, we found that two setups ended with only few haloes above the mass-cut, so we removed them from further analyses.   }

\subsection{$N_{\texttt s}-M$ relation fit}
\label{sec:nsf}

\begin{figure*}
   \includegraphics[width=\textwidth]{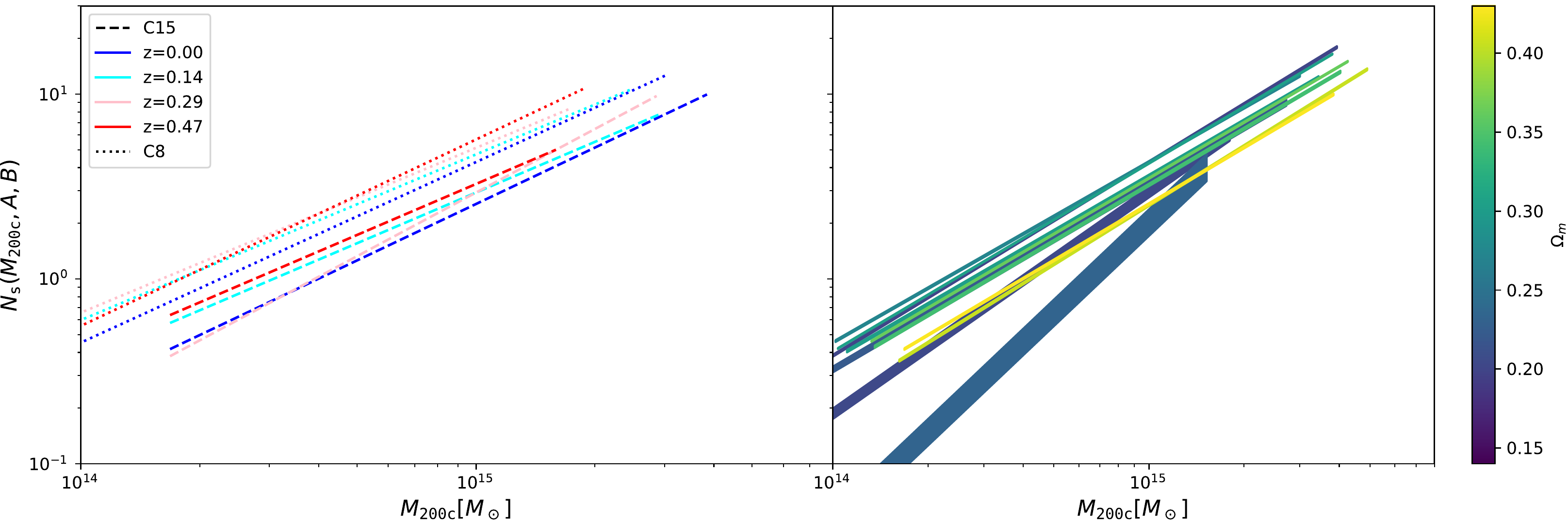}
   \caption{Average satellite count 
   within $R_\mathrm{200c}$
   $vs.$ 
   halo mass for different simulations and redshifts as resulting from maximising Likelihood in Eq. \eqref{eq:l}. Left panel: relation for simulations Box1a/mr  C8 (dashed lines) and C15(dotted lines) at 3 different redshifts (the redder the redshifted). Right panel: each line represents a simulation at $z=0,$ color coded with green with increasing $\Omega_m;$ line tickness covers the gaussian scatter (poissonian scatter is omitted).
   }
    \label{fig:ns_sims_overview}%
\end{figure*}

\begin{figure*}
   \centering
    \includegraphics[width=\textwidth]{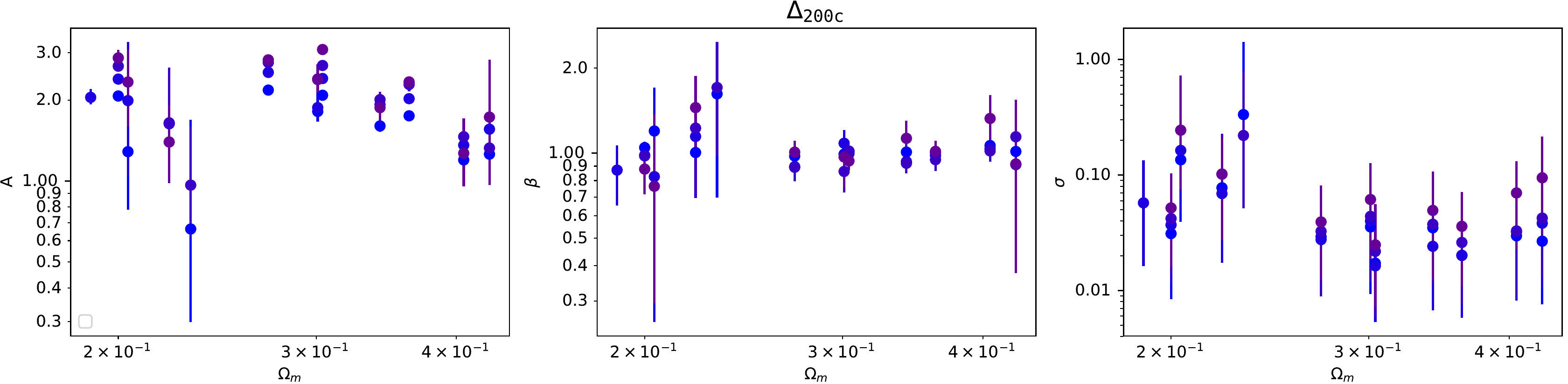}
   \caption{Fit parameters of Eqs. \eqref{eq:l} and \eqref{eq:Ns} for $N_\texttt{s,200c}.$  
   From left to right, parameters $A, B, \sigma$ as a function of $\Omega_m$ and colour-coded by $1+z$ (the redder, the higher the red-shift).
   Vertical error bars corresponds to the uncertainty given from the Likelihood posterior in Eq.  \eqref{eq:l}.}
   \label{fig:hod_minitabella_abs}%
\end{figure*}


\agg{We model the average satellite abundance as a power law of halo mass as in Eq. \eqref{eq:Nsk}, with a normalisation $A$ and log-slope $\beta$ as follows}
\begin{equation}
\langle N_s \rangle_M  =
N_\texttt{s} \left(M, A, \beta\right) 
=  A \cdot \left(\frac{M}{M_p}\right)^\beta,
\label{eq:Ns}
 \end{equation}
{where we use  $M_p=5\times10^{14}M_\odot$ as pivot mass, because it approximates the median mass of the haloes selected  at $z=0$ in the reference cosmology $C8.$}
 
 \agg{We fit $A$ and $\beta$ parameters by maximising a likelihood $\mathcal{L}$  that models satellite HOD as the convolution of a Poisson and a positive-value Gaussian distribution, with fractional scatter $\sigma$ on the average satellite abundance.
 This kind of modelling has been used in the mass-calibration studies as \cite{2019MNRAS.488.4779Costanzi} and \cite{2020PhRvD.102b3509Abbott} where the positive-values Gaussian scatter accounts for different accretion histories. The likelihood results as follows:}  
\begin{equation}
 \mathcal{L}\left(A, \beta, \sigma\right)
 =
 \prod_i {
    \frac{
        \int_0{
           \mathcal{P}\left(  N_{\texttt s,i} | n \right)
            \mathcal{N}\left( 
                n
                 \big| 
                  N_\texttt{s} \left(M_i, A, \beta\right),
                 \sigma N_\texttt{s} \left(M_i, A, \beta\right)
            \right)          }
     \mathrm{d}n
    }
    {
        \int_0{
            \mathcal{N}\left( 
                n
                 \big| 
                  N_\texttt{s} \left(M_i, A, \beta\right),
                 \sigma N_\texttt{s} \left(M_i, A, \beta\right)
            \right)  
        }
     }
 } \, ,
 \label{eq:l}
\end{equation}
where $i$ runs over {all haloes that we selected in a snapshot}.

We maximised\footnote{We used \texttt{python} package \texttt{emcee}  from \cite{2019zndo...3543502Foreman-Mackey} 
}
the likelihood in Eq. \eqref{eq:l} for all simulations separately and Figure \ref{fig:ns_sims_overview}  shows the power law fit for C8 and C15 at all available redshift  (left panel) and for all simulations at $z=0$ (right panel). {The shaded area corresponds to} the Gaussian scatter $\sigma$, showing that average satellite abundances differ on different cosmologies.  
{We can qualitatively see that different cosmologies and redshifts lead to values of $\beta$ that are close to $1$ and a normalisation that can vary up to a factor of two.
 See Appendix \ref{ap:fit_vir_200c} for more details on the fit values.}

One may argue that the dependency from cosmological parameters on $N_s$ is rooted on the fact that $N_s$  depends on halo concentration and the halo concentration depends on cosmology.
However, even if concentration and assembly bias  play a role in the satellite count, it cannot account completely for the dependency between cosmological parameters shown in Fig. \ref{fig:ns_sims_overview}. In fact, Box1a/mr C13 simulation has an outstandingly low  number of satellites (it has no halo with $N_s\geq8$ and in fact it is not included in the emulator), while it has not a particularly high concentration-mass  normalisation \citep[see Fig. 2 in ][]{2021MNRAS.500.5056Ragagnin}.
\agg{In Figure \ref{fig:hod_minitabella_abs}, we summarise the parameters $A, \beta,$ and $\sigma$  found by maximising Eq. \eqref{eq:l} for 
$\Delta_\texttt{200c}$, where we can see a mild redshift evolution of $A$ and $\beta$ as found in \cite{2004ApJ...609...35Kravtsov}.}
{An increase of $N_s$  with redshift could be expected since at high redshift we are selecting more and more young clusters (the mass-cut does not change with redshift), and young clusters are known to be richer ~\citep[see][]{2019MNRAS.490.5693Bose}.}

\subsection{The gaussian process regression emulator}
\label{sec:emu}

In order to model the HOD as a function of  cosmological parameters and redshift, we will build  an emulator based on Gaussian process regression (GPR) with the aim of predicting  $A,\beta,$ and $\sigma.$
Our main motivation is that these parameters 
do not follow simple functional forms, as for instance a power law, as can be seen in Fig. \ref{fig:hod_minitabella_abs}.

For this purpose we train the GPR model\footnote{We used \texttt{sklearn} package \citep{scikit-learn}.} on an array of $A_i, \beta_i,$ and $\sigma_i$   residuals with respect to a power-law fit on cosmological parameters (where $i$ runs on all setup). We present fit posteriors in Appendix \ref{ap:posterior}, and here below we report the results and errros from the fit:
\begin{equation}
\begin{split}
\textrm{ln}\left(A_{\texttt 200c}\right) = & 
 0.551_{-0.041}^{0.045}
 -1.304_{-0.224}^{0.246}\textrm{ln}\Bigl(\frac{\Omega_m}{\Omega_{m,p}}\Bigr) +3.008_{-1.104}^{1.093}\textrm{ln}\Bigl(\frac{\Omega_b}{\Omega_{b,p}}\Bigr)
\\
&+4.037_{-0.823}^{0.610}\textrm{ln}\Bigl(\frac{\sigma_8}{\sigma_{8,p}}\Bigr)
 -0.803_{-1.982}^{1.974}\textrm{ln}\Bigl(\frac{h_0}{h_{0,p}}\Bigr)
 \end{split}
\label{eq:A200c}
\end{equation}
\begin{equation}
\begin{split}
&-0.878_{-0.323}^{0.267}\textrm{ln}\Bigl(\frac{a}{a_p}\Bigr)
\pm0.2\\
\textrm{ln}\left(\beta_{\texttt 200c}\right) =& 
 0.043_{-0.030}^{0.028}
 +0.288_{-0.185}^{0.177}\textrm{ln}\Bigl(\frac{\Omega_m}{\Omega_{m,p}}\Bigr)
-0.931_{-0.826}^{0.850}\textrm{ln}\Bigl(\frac{\Omega_b}{\Omega_{b,p}}\Bigr)
\\
& -1.056_{-0.664}^{0.690}\textrm{ln}\Bigl(\frac{\sigma_8}{\sigma_{8,p}}\Bigr)  -0.775_{-1.586}^{1.590}\textrm{ln}\Bigl(\frac{h_0}{h_{0,p}}\Bigr)
\\&+ 0.080_{-0.155}^{0.162}\textrm{ln}\Bigl(\frac{a}{a_p}\Bigr)
\pm0.1,
\end{split}
\label{eq:B200c}
\end{equation}
where pivot cosmology parameters  are set to C8 values and pivot scale factor is $a=0.87.$

We trained our emulator on  log-scaled values, as follows:

\begin{equation}
\begin{cases}
X_i=\Bigl[
    
        \textrm{ln}\Bigl(\dfrac{\Omega_{m,i}}{\Omega_{m,p}}\Bigr),
        \textrm{ln}\Bigl(\dfrac{\Omega_{b,i}}{\Omega_{b,p}}\Bigr),
        \textrm{ln}\Bigl(\dfrac{\sigma_{8,i}}{\sigma_{8,p}}\Bigr),
        \textrm{ln}\Bigl(\dfrac{h_{0,i}}{h_{0,p}}\Bigr),
        \textrm{ln}\Bigl(\dfrac{1+z_p}{1+z_i}\Bigr)
    \Bigr]\\
    \\
y_i=\Bigl[
        \textrm{ln}\Bigl(\dfrac{A_i}{A_\Delta}\Bigr),
        \textrm{ln}\Bigl(\dfrac{\beta_i}{\beta_\Delta}\Bigr),
        \textrm{ln}\Bigl(\dfrac{\sigma_i}{\sigma_\Delta}\Bigr)
    \Bigr]\ ,
\end{cases}
\end{equation}
where  $X=\left\{X_i\right\}$ is the input data; $y=\left\{y_i\right\}$ the output data; $i$ runs over all data points (i.e. all selected snapshots) for which we maximised Likelihood in Eq. \eqref{eq:l},  and $A_\Delta, \beta_\Delta,$ and $\sigma_\Delta$ are a function of cosmology and, as  pivot values, we used  the same as in Eq. \eqref{eq:A200c} and Eq. \eqref{eq:B200c}.

Concerning the GPR model, we modelled our kernel $K$ as a constant $K_0,$ times a gaussian  Radial-basis function (RBF) kernel with length scale $l$:
\begin{equation}
    K(\mathbf{x_1},\mathbf{x_2}) = K_0 \times \mathrm{exp}\left(-\frac{\left\lVert\mathbf{x_1}-\mathbf{x_2}\right\lVert^2}{2l^2}\right),
\end{equation}
where the norm $\left\lVert\dotsc\right\rVert$ is the euclidean distance. 

We maximised the log marginal likelihood as proposed in Eq. 2.30 in \cite{rasmussen}   and let  parameters $K_0$ and $l$ to vary in the maximisation.

Hereafter we define the Emulator predictions as $A^\textrm{Emu},$ $\beta^\textrm{Emu},$ which themselves  depends on a  cosmology and scale factor $(\Omega_m, \Omega_b, \sigma_8, h_0,a).$
We define the emulated average number of satellites $N_{\mathrm {s,Emu}}$ as
\begin{equation}
 N_{\textrm{s,Emu}}  = A_\textrm{Emu} \times 
\left(\frac{M}{M_p}\right)^{ {\beta_\textrm{Emu}}},
\label{eq:NsEmu}
\end{equation}
{where  $A_\textrm{Emu},$ and $\beta_\textrm{Emu}$ are predicted by our emulator an depend on
 cosmology  and redshift.}

\subsection{Emulator error estimate}
\label{sec:er}

\begin{figure}
  \centering
   \includegraphics[width=.35\textwidth]{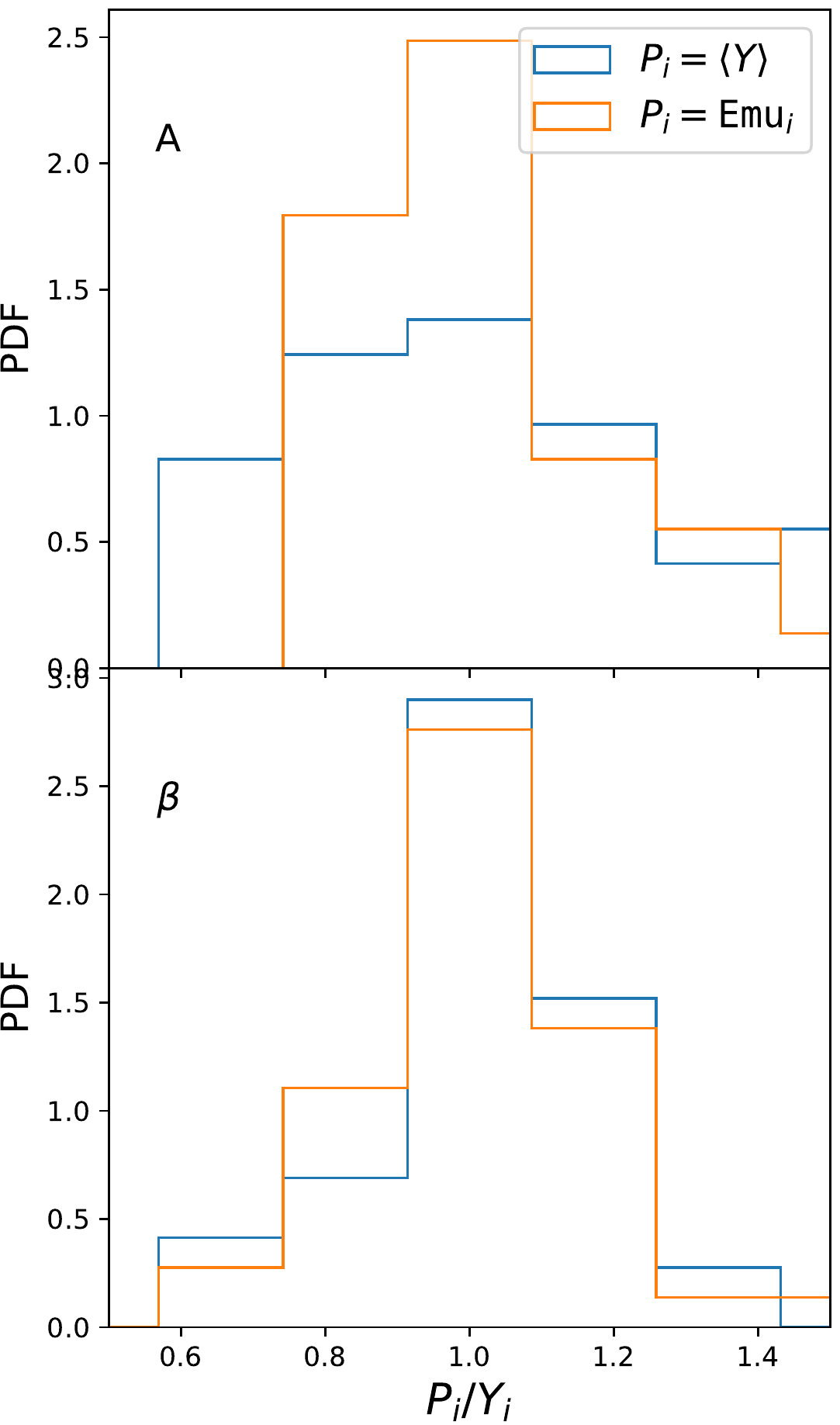}
   \caption{PDF of residuals for predictor $P_i=\textrm{Emu}_i$ and the residual based on predictions from the average values ($P_i=\langle Y\rangle$), with respect to the missing data point  $y_i.$
   Data is computed on overdensity $\Delta_\mathrm{200c}.$
   Upper row show PDF for $y_i=A_i,$ lower row show PDF for $y_i=\beta_i.$ The scatter obtained with the emulator is significantly smaller than the residuals on the averages}
    \label{fig:emu_sebastian_residuals}%
\end{figure}

To estimate the precision of our emulator  we use the same technique as used by \citet{2020ApJ...901....5Bocquet}: for each data vector available $(X_i, y_i)$, we (i) build a predictor trained on the complete data-set but that point (i.e. $[X,Y]_i = [\{X\}\setminus X_i, \{Y\}\setminus y_i]$), hereafter $O_{{\tt Emu},i}$ and (ii) for each predictor we compute its relative error in predicting the un-trained value $y_i.$ 

To contextualise the relative error of the emulator, we will  compare it with the relative error  obtained by predicting $y_i$ using only the average of all values (thus ignoring any cosmology dependency).
 
As we can see in Figure \ref{fig:emu_sebastian_residuals}, the residual PDF {of the $A$ from the emulator emulators (top panel, orange steps)} is much more peaked around $1$  respect to the PDF from the predictor based on the averages (blue steps).
This implies  that the emulator is  effective in predicting mass-richness normalisation $A$.
{On the other hand there is no significant gains in recovering the log-slope $\beta.$}

\agg{The residuals distribution of the emulator within $\Delta_\mathrm{200c}$ corresponds to a precision of $\approx10-20\%,$  and the average of the  GPR error estimations in the missing points is of the same order of magnitude, thus  the emulator is capable of correctly predicting  its own uncertainty.}

\subsection{Mass range}

\begin{figure*}
   \centering
   \includegraphics[width=0.9\textwidth]{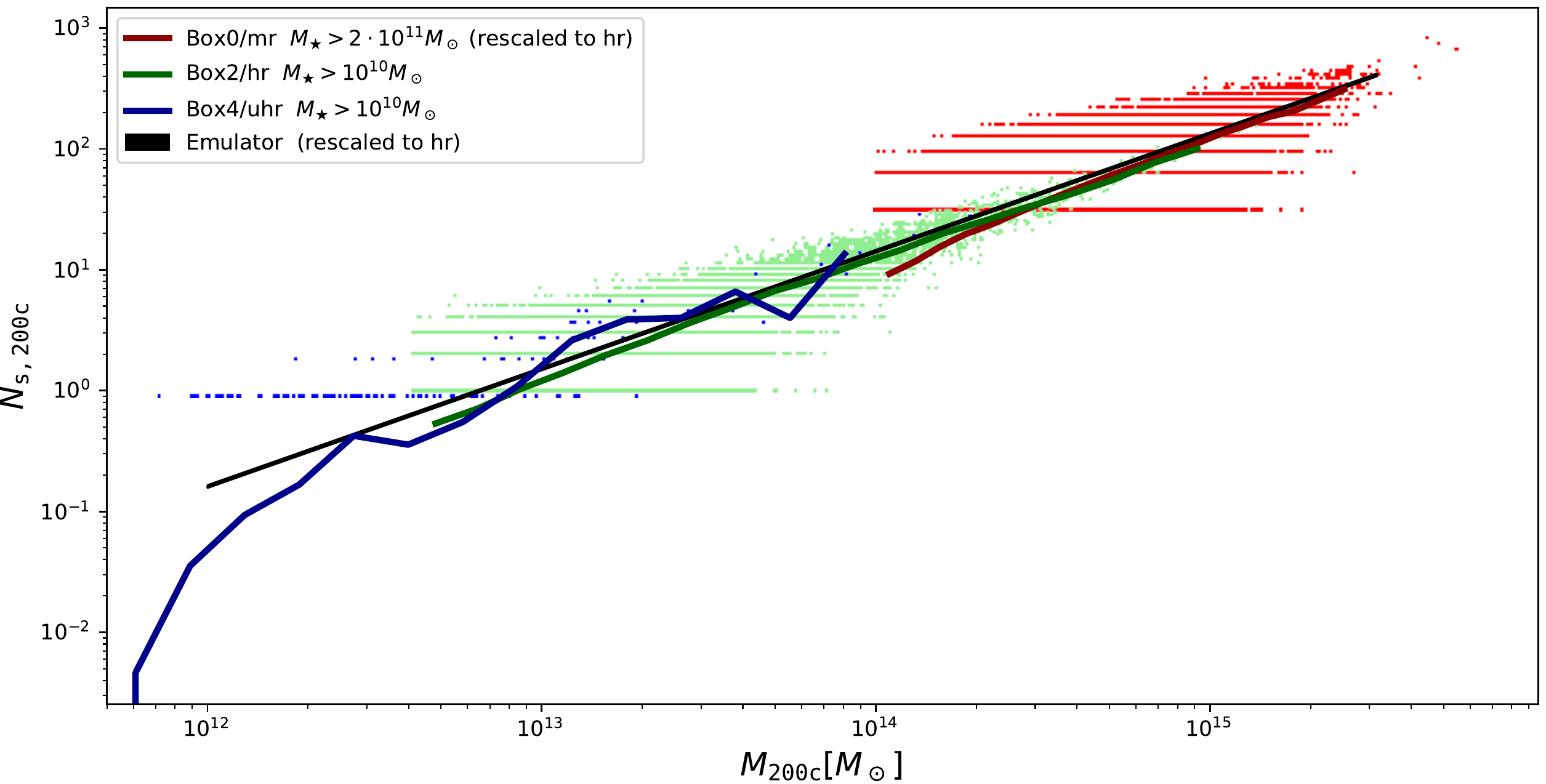}
   \caption{Satellite count $N_{\texttt s, 200c}$  $vs.$ halo mass $M_{\texttt 200c}$ for three Magneticum simulations, Box4/uhr (blue coloured),  Box2/hr (green coloured), and  Box0/mr (red colour) to account for resolution effects. Data points represents single haloes, coloured lines represents average values per mass bin. Black line is emulator prediction.
   Emulator and \texttt{Box0/mr} data are re-scaled with Eq. \eqref{eq:Nsr}.}
    \label{fig:massrange}%
\end{figure*}

In this subsection we test the mass range validity of our satellite HOD across various orders of magnitude in order to identify the halo mass range of it.

Figure \ref{fig:massrange} shows that the same power law satellite abundance holds from the most massive galaxy clusters $M_{200c}\approx6\cdot10^{15}\Msun$ down to haloes of $M_{200c}\approx5\cdot10^{13}\Msun.$ At that mass there starts the low-mass drop of $M_\star>10^{10}\Msun$ cut, which is particularly visible in the Box4/uhr regime haloes.
Both Box0/mr and C8 satellite abundances are re-scaled using in Eq. \eqref{eq:Nsr}.
{The match between mass-richness relations from our emulator and   various Magneticum boxes (some of which are rescaled using Equation \ref{eq:Nsr}), shows that Equation \eqref{eq:Nsr} is consistent between resolutions and that our relatively high stellar-mass cut is enough to constrain the normalisation and log-slope of the mass-richness relation.} 

\subsection{Comparison with numerical studies}
\label{sec:compare}


\begin{figure}
   \centering
   \includegraphics[width=.43\textwidth]{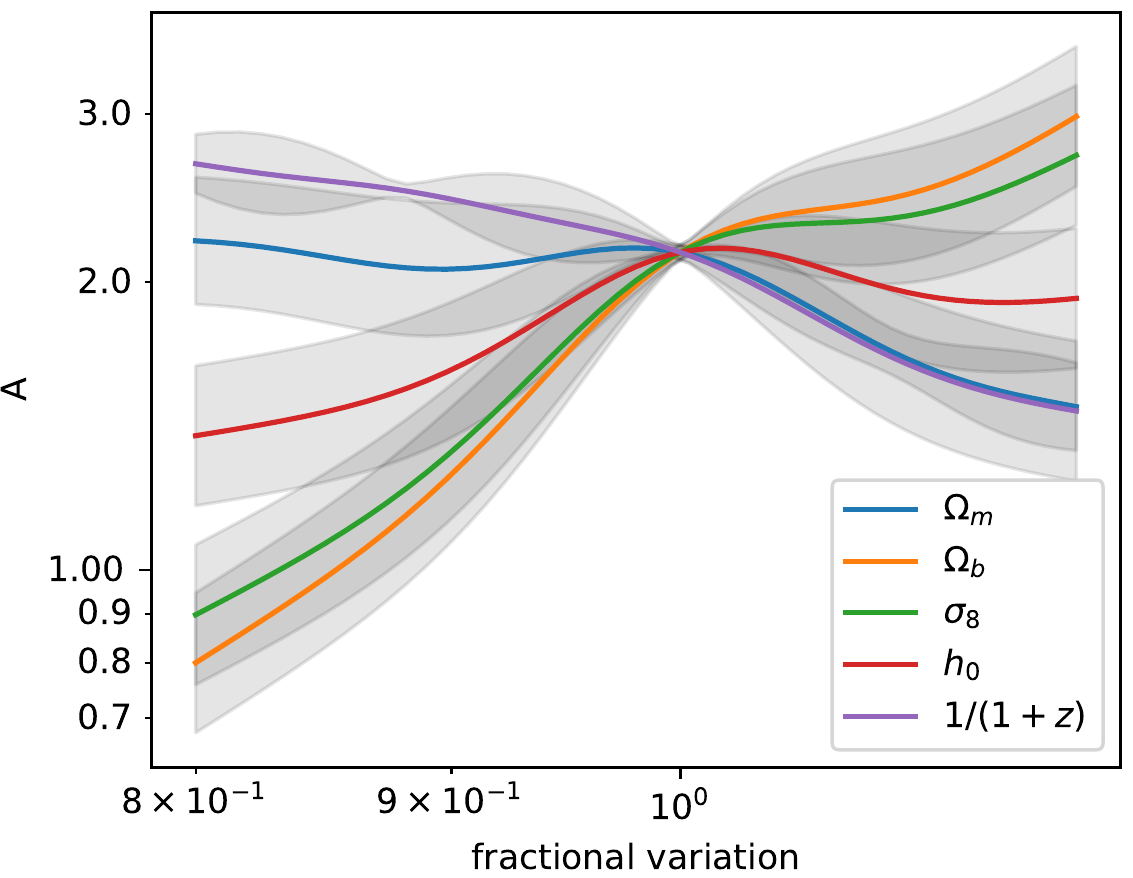}
   \caption{Variation of  $N_\mathrm{s,200c}$ emulator    $A$,  $\beta$ and   $\sigma$ as a function of fractional variation of cosmological parameters $\Omega_m, \Omega_b, \sigma_8, h_0$ and scale factor. Shaded areas show $1$ standard deviation provided by the Gaussian Process error estimation. 
   }
    \label{fig:emu_variation}%
\end{figure}

We now investigate the effect of each cosmological parameter  on the   $N_s-M$ relation.
Figure \ref{fig:emu_variation} shows the parameters $A_{\tt Emu}, B_{\tt Emu}$ variation from WMAP7 cosmology values, as a function of fractional variation of each cosmological parameter $\Omega_m, \Omega_b, \sigma_8, h_0,$ and $a$ separately.

The normalisation decreases with scale factor, which is in agreement with the fact that BAHAMAS (that runs at $z=0.1$) has a higher normalisation than Magneticum~\citep{2020MNRAS.495..686Anbajagane}, yet it has very similar  cosmological parameters. 
{Note that  this manuscript do not aim to predict HODs of the other simulation suites, rather, we show that their variation is comparable with the one predicted by changing cosmological parameters alone on  fixed-feedback simulations.
In Appendix \ref{ap:compa} we compare in detail the mass-richness relation of our emulator and the one of other hydrodynamic simulations in the literature.}

\section{Impact on mock observations}
\label{sec:mock}
In this section we test the cosmology-dependence of the HOD on mock catalogues, in order to estimate the impact of such dependence on the cosmological parameter constraints. To this purpose, we consider the richness, a weighted sum of the galaxy members, often used as mass proxy in photometric cluster surveys. {We recast Eq. \eqref{eq:Ns} in terms of a richness-mass relation:}
\begin{equation}
    \langle \lambda \rangle_M  =  A_\lambda \cdot \left(\frac{M}{M_p}\right)^{\beta_\lambda},
    \label{eq:lambda}
\end{equation}
with $A_\lambda = A_0 \cdot A_{\rm emu} = 72.4 \pm 0.7$ and $\beta_\lambda = \beta_0 \cdot \beta_{\rm emu} = 0.935 \pm 0.038$ \citep[from table IV of][]{DES:2020cbm}. Here, $A_{\rm emu}$ and $\beta_{\rm emu}$ are the predictions of the emulator and contain the dependence on cosmology, while $A_0$ and $\beta_0$ represent the cosmology-independent part of the total parameters. 

To perform the analysis, we extract a catalogue of halo masses corresponding to the C8 simulation at redshift $z=0$, following the \citet{Despali:2015yla} analytical mass function. This step ensures to have a proper description of the mass function, in order to obtain an unbiased estimation of parameters. We obtain a catalogue with $\sim 2.8 \cdot 10^5$ objects, with virial masses above $M_\mathrm{vir}>10^{13}\Msun$, to which we assign richness by applying Eq. \eqref{eq:lambda}, plus a Poisson  scatter. To ease the analysis, we neglect the intrinsic scatter of the HOD, which is subdominant with respect the Poisson  one. In the end, we compute the number counts by considering 5 richness bins, between $\lambda = 30 - 300$, where the sample is complete in mass.

Then, {we maximise a Gaussian likelihood}
\begin{equation}
    \label{gauss_cov_l}
        \mathcal{L}(x\,\vert\,\mu,\,C) = \ffrac{ \exp {\left \{-\frac{1}{2} (\mathbf{x}-\boldsymbol{\mu})^T C^{-1}  (\mathbf{x}-\boldsymbol{\mu}) \right \}}}{\sqrt{2\pi \det C}} \,,
\end{equation}
{where $\mathbf{x}$ is the mock "observed" number count and $\boldsymbol{\mu}$ is a MCMC test one},  and $C$ is the covariance matrix, computed following the analytical model of \citet{Hu:2002we}. Since in this test we only aim to give an estimation of the impact of the cosmology-dependent HOD, we run a simplified Monte Carlo Markov Chain (MCMC) process with only two free cosmological parameters, $\Omega_{\rm m}$ and $\log_{10} A_s$ (and thus \s8), neglecting the dependence of the HOD on $\Omega_{\rm b}$ and $h_0$, and neglecting the redshift dependence. 

Following the approach of \citet{Singh:2019end}, we compare three different cases:
\begin{enumerate}[i]
    \item \textit{no cosmo} case: we ignore the cosmology dependence of the HOD, so that  $A_\lambda = A_0$ and $\beta_\lambda = \beta_0$. We assume flat uninformative priors both on \om and \logas and on $A_\lambda$ and $\beta_\lambda$;
    \item \textit{cosmo} case: we assume flat uninformative priors on \om, \logas, $A_0$ and $\beta_0$, plus Gaussian priors on $A_\lambda$ and $\beta_\lambda$, respectively given by $\mathcal{N}(72.4, 7.0)$ and $\mathcal{N}(0.935, 0.038)$. The cosmology-dependent parameters $A_{\rm emu}$ and $\beta_{\rm emu}$ are computed by the emulator at each step of the MCMC process, and, to take into account the emulator inaccuracy, we randomly extract a value from a Gaussian distribution with center in the emulator prediction and amplitude equal to $\sigma_{\log A_{\rm emu}} = 0.06$ and $\sigma_{\log \beta_{\rm emu}} = 0.09$;
    \item \textit{cosmo + WL} case: we add the weak lensing (WL) cosmological dependence which affects the mass calibration in the real observations, to figure out whether the combination of the cosmology-dependent HOD with other cosmological probes could improve the parameter constraints. We model such dependence by modifying the prior on $A_\lambda$, which becomes a Gaussian prior with the same amplitude of the previous case, but ceneterd on
    \begin{equation}
        A'_\lambda = A_\lambda - \ln 10^{\Delta (\Omega_{\rm m})}
    \end{equation}
    with $\Delta (\Omega_{\rm m}) = \beta_\lambda \, \frac{{\rm d} \ln M_{\rm WL}}{{\rm d} \, \Omega_{\rm m}} (\Omega_{\rm m} - 0.3)$, where $\frac{{\rm d} \ln M_{\rm WL}}{{\rm d} \Omega_{\rm m}} = -0.68$ is the average value from table I of \citet{DES:2018crd}. 
\end{enumerate}

\begin{figure*}
   \centering
   \includegraphics[scale=0.63]{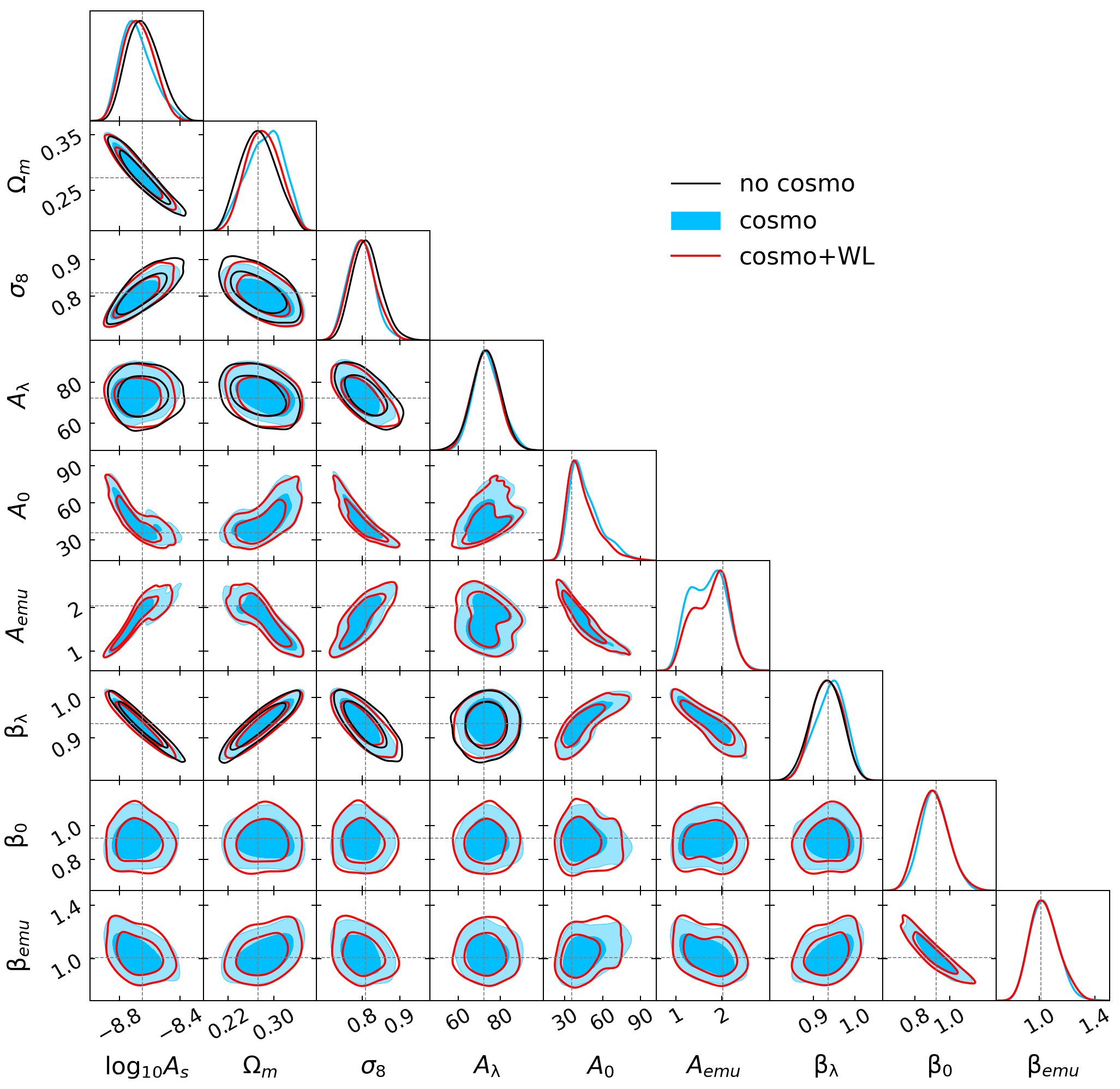}
   \caption{Contour plots at 68 and 95 per cent of confidence level for the three cases described in Sect. \ref{sec:mock}: \textit{no cosmo} (black), \textit{cosmo} (blue) and \textit{cosmo+WL} (red) contours. The grey dashed lines represent the input values of parameters.}
    \label{fig:post_mock}%
\end{figure*}

In Fig. \ref{fig:post_mock}, we show the posterior distributions resulting from the three analysis. As expected, the marginalised posteriors recovered by the \textit{cosmo} case are similar to the ones from the \textit{no cosmo} case, but in addition the former is able to constrain the cosmology-dependent and cosmology-independent components of the richness-mass relation separately. This can represent an advantage, since the components of $A_\lambda$ show stronger degeneracies with cosmological parameters with respect to the one of their combination; such degeneracies can be exploited when combined with other cosmological probes. On the contrary, this decomposition for $\beta_\lambda$ does not present the same advantage, as the full parameter has an higher degeneracy with cosmological parameters with respects to its components.

The third case presents similar posteriors to the simple \textit{cosmo} case; to better compare the differences, we quantify the accuracy of the parameter estimation by computing the figure of merit \citep[FoM,][]{Albrecht:2006um}
\begin{equation}
    {\rm FoM}(\Omega_{\rm m},\sigma_8) = \ffrac{1}{\sqrt{{\rm det} \; C(\Omega_{\rm m},\sigma_8)}}
\end{equation}
where $C(\Omega_{\rm m},\sigma_8)$ is the parameter covariance matrix obtained from the posteriors. The FoM is proportional to the inverse of the area enclosed by the ellipse
representing the 68 percent confidence level: the  higher the FoM, the more accurate is the parameter evaluation. The result, shown in table \ref{table:3}, indicates that the use of the cosmology-dependent HOD allows us to obtain more constraining posteriors, further improved with the addition of the weak lensing information. To prove that the \textit{cosmo+WL} result is not achieved only thanks to the addition of WL, we show also the FoM for the \textit{no cosmo + WL} case, which has a constraining power similar to the simple \textit{no cosmo} case. By comparing the FoM of the three cases, we obtain an improvement of about the 6 percent for the \textit{cosmo} case and of about the 11 percent  for the \textit{cosmo + WL} case with respect the \textit{no cosmo} one.

\begin{table}
\caption{Figure of merit in the \om -- \s8 plane for the three cases of Fig. \ref{fig:post_mock}, plus the \textit{no cosmo + WL} case. In the right column, normalised differences with respect to the \textit{no cosmo} case.}              
\label{table:3}      
\centering                                      
\begin{tabular}{l c c}          
\hline\hline   
Case & FoM & $\Delta$FoM/FoM$_{nc}$  \\
\hline
no cosmo       & 980  & -     \\
no cosmo + WL  & 993  & 0.01 \\
cosmo          & 1044 & 0.06  \\
cosmo + WL     & 1088 & 0.11  \\
\hline
\end{tabular}
\end{table}

\section{Conclusions}
\label{sec:conclu}

We tackled the problem of studying the satellite count $vs.$ halo mass relationship dependency on cosmological parameters and redshift and studied how {such} modelling {could improve observational studies as mass-calibrations that use mass-richness relations}.
To this end we used  FP \magneticum simulations Box1a C1--C15 (see Table \ref{table:1}) that were run with the same initial conditions but different cosmological parameters  $\Omega_m, \Omega_b, \sigma_8, h_0.$
\agg{We did not re-calibrate feedback parameters over the various runs, and in this study we focus on the   effect of cosmological parameters on $N_s$  for a fixed feedback configuration.}
In particular:
\begin{itemize}
    \item We showed that {  DMO and FP subhaloes depends differently from cosmological parameters, showing the importance of parametrising mass-richness relations with results from FP simulations rather than DMO ones.
}
    \item {\bf
     
    We} built an emulator capable of predicting normalisation, log-slope and log-scatter of the high-mass power-law regime of the $N-M$ relation based on GPR modelling, in order to predict the number of satellites for a given cosmology. We estimated its error being within $\approx20\%.$ This error is comparable with the same uncertainty predicted by the GPR, which ensures that the error estimation of the GPR is under control.
   \item 
    {We tested whether parametrising  mass-richness relation with cosmological parameters can improve mass-calibration studies.}
    The likelihood analysis on mock {mass-calibration} showed that the use of a cosmology-dependent HOD provides an improvement ($\sim$ 5 \%) in the constraining power over a simple cosmology-independent HOD, which can be further improved ($\sim$ 10 \%) if combined with multiple mass proxies, such as the weak lensing signal.
\end{itemize}

The study was carried out over a small number of cosmologies and with a resolution limited to the high-mass regime of haloes, {and showed that mass-calibrations can benefit from modelling mass-richness relations with cosmological parameters from hydro-simulations. 
Future studies could focus on the dependency from cosmology of  the radial distribution of substructures.  }
The emulator log-slope predictions have a large uncertainty (see Figure \ref{fig:emu_variation} where $\beta$ shaded area spans between $\beta\in[0.9,1]$) which means there is the need for simulations over additional cosmological parameters \agg{and feedback parameters} in order to improve GPR interpolation.

\begin{acknowledgements}
The \textit{Magneticum Pathfinder} simulations were partially performed at the Leibniz-Rechenzentrum with CPU time assigned to the Project `pr86re'. AF would like to thank Stefano Borgani for useful discussions. AR is supported by the EuroEXA project (grant no. 754337). KD acknowledges support by DAAD contract number 57396842. AR acknowledges support by  MIUR-DAAD contract number 34843  ,,The Universe in a Box''. AS, AF and PS are supported by the ERC-StG 'ClustersXCosmo' grant agreement 716762. AS is supported by the the FARE-MIUR grant 'ClustersXEuclid' R165SBKTMA and  INFN InDark Grant. This work was supported by the Deutsche Forschungsgemeinschaft (DFG, German  Research  Foundation)  under  Germany's  Excellence Strategy - EXC-2094 - 390783311. KD acknowledges funding for the COMPLEX project from the European Research Council (ERC) under the European
Union’s Horizon 2020 research and innovation program grant agreement ERC-2019-AdG 860744. AR acknowledges support from the grant
PRIN-MIUR 2017 WSCC32. 
We are especially grateful for the support by M. Petkova through the Computational Center for Particle and Astrophysics (C$^2$PAP). Information on the \textit{Magneticum Pathfinder} project is available at \url{http://www.magneticum.org}. 
\end{acknowledgements}

%
\bibliographystyle{aau} 
 \bibliography{biblio} 
%

\begin{appendix} 

\section{Stellar mass cut}
\label{ap:conv}

\begin{figure*}
   \centering
   \includegraphics[width=\textwidth]{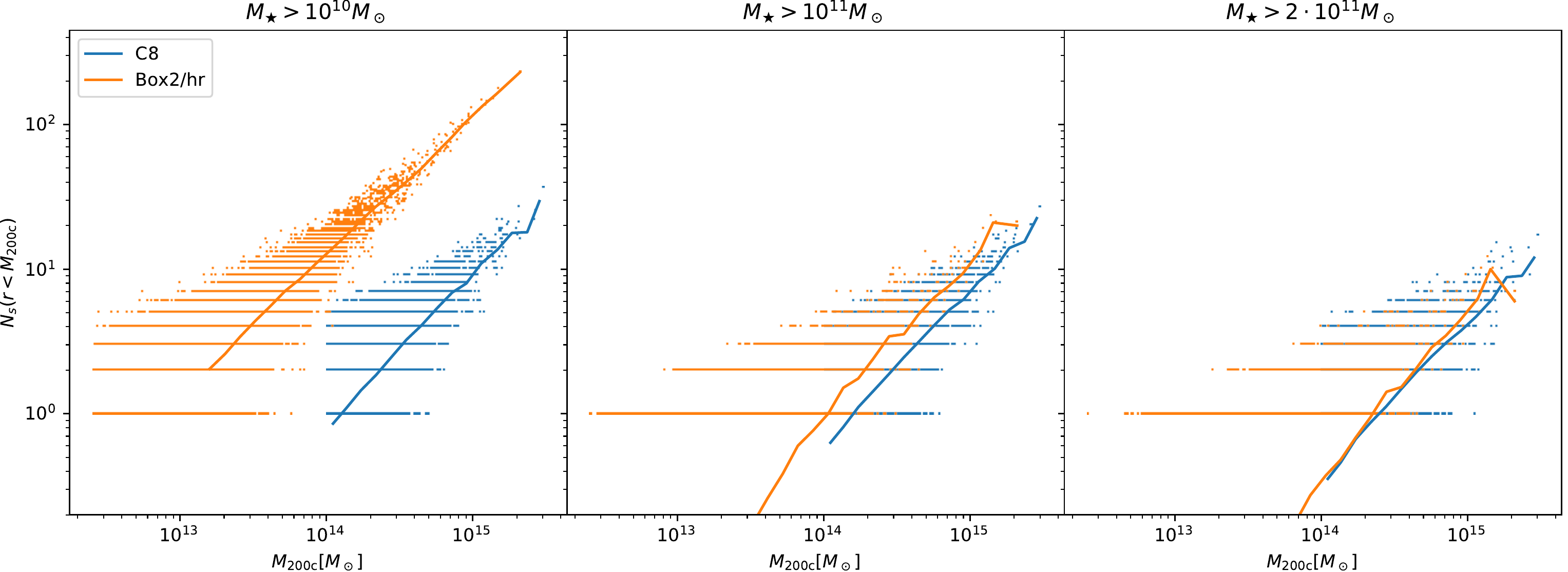}
   \caption{Galaxy abundance of HR (\texttt{Box2/hr} in orange) and MR (C8 in blue) simulations for varying minimum stellar mass of galaxies: $M_\star>10^{10}\Msun$ (left panel), $M_\star>10^{11}\Msun$ (middle panel), an  $M_\star>2\cdot10^{10}\Msun$ (right panel).}
    \label{fig:hod_c8_vs_box2bhr}%
\end{figure*}

To build our HOD emulator, first of all we estimate which stellar mass cut to apply to the satellite count of our Box1a/mr C1--C15 simulations.

For this purpose we vary the stellar mass cut of both C8 and its HR counterpart (Box2/hr)  until their $N_s-M$ relations statistically match. 
Figure \ref{fig:hod_c8_vs_box2bhr} show convergence tests for stellar mass of C8 satellites.
Left panel shows the $Ns-M$ relation of C8 and Box2/hr with the fiducial Box2/hr mass-cut $M_\star>10^{10}\Msun$ \citep[as chosen in][]{2020MNRAS.495..686Anbajagane}; central panel shows that a stellar mass cut $M_\star>10^{11}\Msun$ is not high enough to make C8 match its HR counter-part; right panel shows that a stellar mass cut $M_\star>2\cdot10^{11}\Msun$ is capable reconciling the two simulations.

\section{Satellite abundance fit for each simulation}
\label{ap:fit_vir_200c}

\begin{table*}
\caption{Each row presents data for a given cosmolgy C1--C15 of Box1a/mr simulations, columns are grouped by redshift slices, and each redshift slice reports the halo mass cut $M_{\rm cut}$ in units of $10^{14}M_\odot$ and the corresponding parameters $A, B$ and $\sigma$   of Eq. \eqref{eq:Ns}. Upper rows present data for overdensity $\Delta_vir$ and lower rows for overdensity 200c.}
\label{table:hod_vir_megatabella}      
\centering                                      
\begin{tabular}{l | rrrr  | rrrr  | rrrr  | rrrr 
}          
\hline\hline
 & \multicolumn{4}{|c|}{$z=0.00$}&\multicolumn{4}{|c|}{$z=0.14$}&\multicolumn{4}{|c|}{$z=0.29$}&\multicolumn{4}{|c}{$z=0.47$}\\
 &  $M_{\rm cut}$ & ln$A$ & $\beta$ & $\sigma$ 
 &  $M_{\rm cut}$ & ln$A$ & $\beta$ & $\sigma$ 
 &  $M_{\rm cut}$ & ln$A$ & $\beta$ & $\sigma$ 
 &  $M_{\rm cut}$ & ln$A$ & $\beta$ & $\sigma$ \\
\hline
C1 & - & - & - & -  & - & - & - & -  & - & - & - & -  & - & - & - & -  \\
C2 & $15.2$ & $-0.76$ & $1.76$ & $1.39$  & $4.0$ & $0.72$ & $0.87$ & $0.06$  & - & - & - & -  & - & - & - & -  \\
C3 & $4.8$ & $0.73$ & $1.05$ & $0.03$  & $4.5$ & $0.87$ & $0.98$ & $0.04$  & $4.5$ & $0.98$ & $0.98$ & $0.04$  & $5.0$ & $1.05$ & $0.88$ & $0.05$  \\
C4 & $9.6$ & $0.25$ & $1.20$ & $0.14$  & $9.7$ & $0.69$ & $0.83$ & $0.16$  & - & - & - & -  & $6.6$ & $0.85$ & $0.76$ & $0.24$  \\
C5 & $8.4$ & $0.49$ & $1.01$ & $0.08$  & $7.1$ & $0.50$ & $1.15$ & $0.07$  & $9.6$ & $0.49$ & $1.23$ & $0.10$  & $7.5$ & $0.33$ & $1.45$ & $0.10$  \\
C6 & $11.0$ & $-0.41$ & $1.62$ & $0.33$  & - & - & - & -  & $7.8$ & $-0.04$ & $1.71$ & $0.22$  & - & - & - & -  \\
C7 & $16.3$ & $-0.88$ & $1.62$ & $1.40$  & $6.0$ & $0.61$ & $0.83$ & $0.07$  & $6.1$ & $0.84$ & $0.72$ & $0.09$  & $5.7$ & $0.84$ & $0.52$ & $0.11$  \\
C8 & $4.3$ & $0.78$ & $0.98$ & $0.03$  & $4.8$ & $0.93$ & $0.90$ & $0.03$  & $5.1$ & $1.01$ & $0.89$ & $0.03$  & $4.6$ & $1.04$ & $1.01$ & $0.04$  \\
C9 & $7.9$ & $0.60$ & $0.99$ & $0.04$  & $8.1$ & $0.63$ & $1.08$ & $0.04$  & $7.1$ & $0.87$ & $0.86$ & $0.04$  & $7.0$ & $0.87$ & $0.97$ & $0.06$  \\
C10 & $4.8$ & $0.73$ & $1.02$ & $0.02$  & $4.0$ & $0.88$ & $1.00$ & $0.02$  & $3.6$ & $0.99$ & $1.02$ & $0.02$  & $4.5$ & $1.13$ & $0.94$ & $0.02$  \\
C11 & $7.4$ & $0.47$ & $1.01$ & $0.03$  & $7.3$ & $0.65$ & $0.93$ & $0.02$  & $5.4$ & $0.70$ & $0.92$ & $0.04$  & $7.3$ & $0.63$ & $1.13$ & $0.05$  \\
C12 & $4.8$ & $0.56$ & $1.00$ & $0.02$  & $6.1$ & $0.70$ & $0.98$ & $0.02$  & $7.1$ & $0.83$ & $0.95$ & $0.03$  & $5.9$ & $0.85$ & $1.02$ & $0.04$  \\
C13 & - & - & - & -  & - & - & - & -  & - & - & - & -  & - & - & - & -  \\
C14 & $10.4$ & $0.18$ & $1.06$ & $0.03$  & $9.4$ & $0.31$ & $1.04$ & $0.03$  & $7.6$ & $0.38$ & $1.02$ & $0.03$  & $10.2$ & $0.24$ & $1.33$ & $0.07$  \\
C15 & $9.4$ & $0.23$ & $1.01$ & $0.03$  & $9.2$ & $0.44$ & $0.91$ & $0.04$  & $9.6$ & $0.28$ & $1.14$ & $0.04$  & $10.5$ & $0.55$ & $0.92$ & $0.09$  \\
\end{tabular}
\end{table*}

Table \ref{table:hod_vir_megatabella} reports the fit parameters of average satellite abundance  a function of halo mass,  for all setups that had a $p-$value below $0.9$ and higher than $0.01.$ The problematic fits were the one at $z=0.67$ where few of them failed, probably because the resolution of our simulations are not always enough to reach this redshift. Only a total of $48$ fits were successful.

In order to fit the power-law halo-mass region of the satellite abundance relation, we imposed a halo mass cut (see Sec. \ref{sec:conv}) at $M_\texttt{vir}=3\cdot10^{14}M_\odot$ for C8 simulation and scaled the mass cut to other simulations according to their baryon fraction.  Some cuts have been modified in order to manually shrink or enlarge the halo range so to maximise the number of data points and yet do not cross the mass cut at low halo masses.

\begin{figure}
   \centering
   \includegraphics[width=0.5\textwidth]{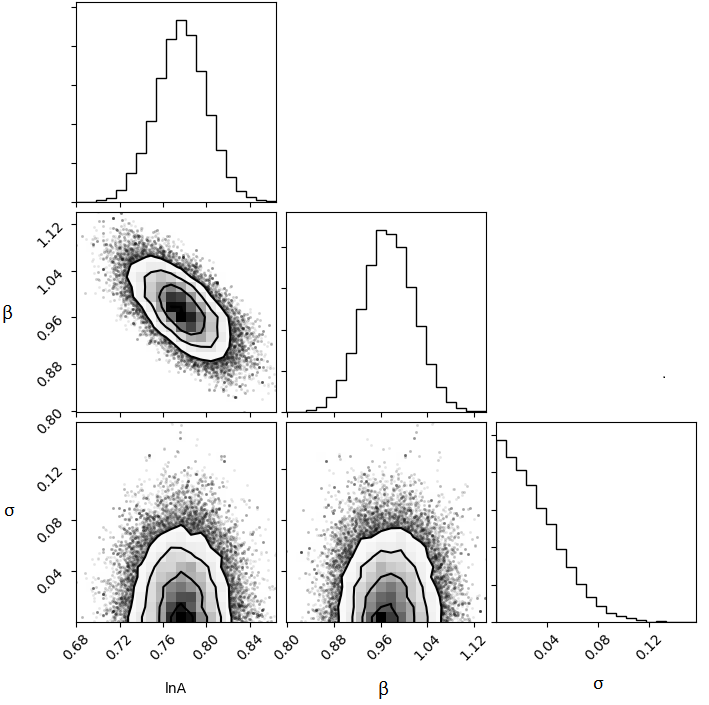}
    \caption{Posterior associated to Likelihood \eqref{eq:l} for Box1a/mr C8 at $z=0$ for $\Delta=$ 200c. Parameters $A$ and $\beta$ are as in Eq. \eqref{eq:Ns} and $\sigma$ is the fractional-scatter in the Likelihood.}
  \label{fig:posteriorC8}%
\end{figure}

Figure \ref{fig:posteriorC8} shows the posterior of the fit for simulation Box1a/mr C8 $z=0.$ Here we can see that the fractional scatter $\sigma$ is consistent with zero.

\section{Cosmology dependent power-law fit posteriors}
\label{ap:posterior}

\begin{figure*}
   \centering
   \includegraphics[width=\textwidth]{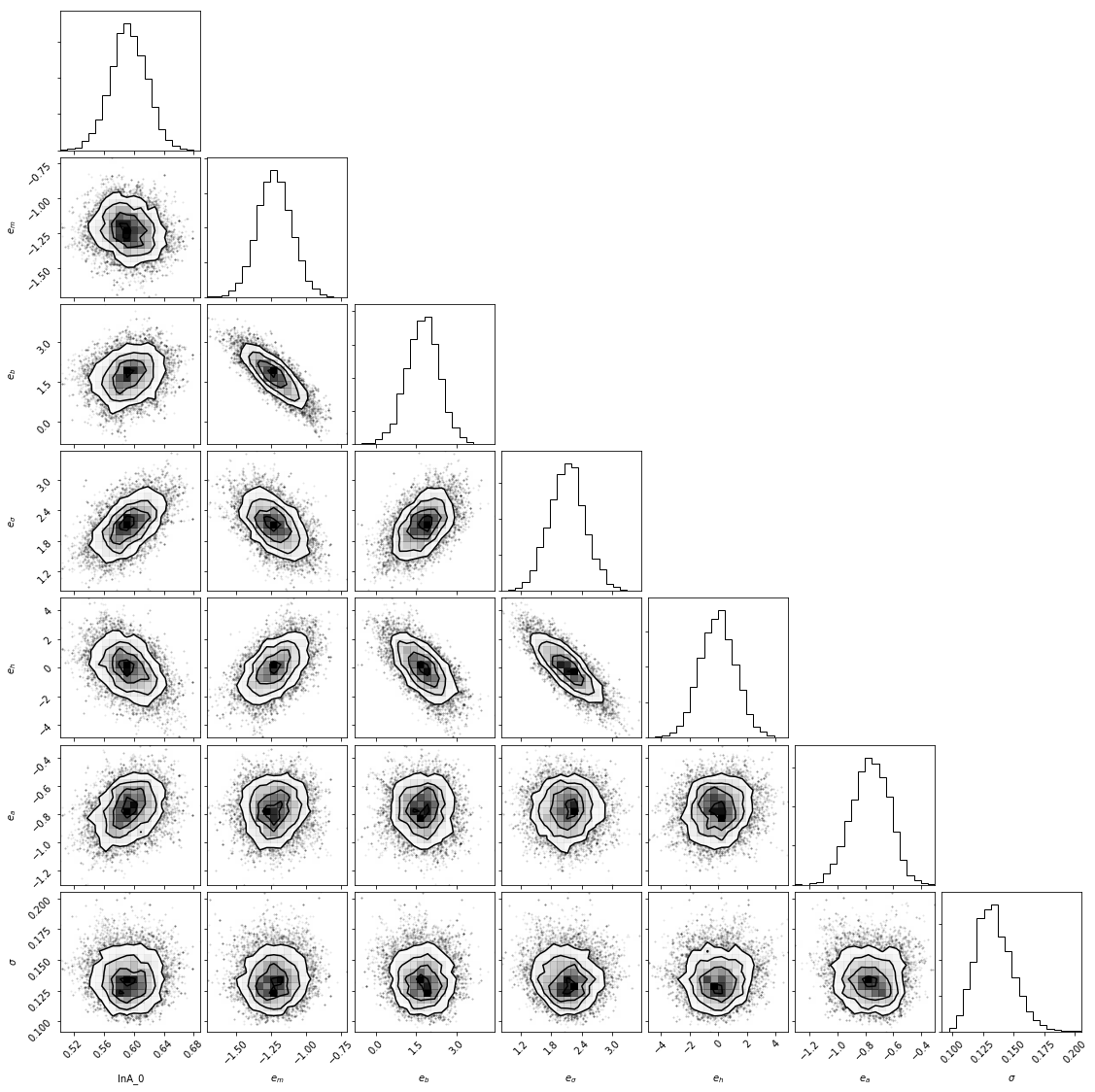}
    \caption{Posterior associated to Likelihood  in Eq. \eqref{eq:l}.
    The parameter $lnA_0$ is the normalisation $0.64_{-0.03}^{+0.03}$ in Eq. \ref{eq:A200c}), 
    $\sigma$ is the gaussian-scatter of the fit, and $e_m, e_b, e_\sigma, e_h, e_a$ are, respectively, the exponents for logarithm of $\Omega_m,\Omega_b,\sigma_8,h_0,1/(1+z)$ divided by the respective pivot as disuccsed in Sec. \ref{sec:nsf}.}
  \label{fig:posteriorA200c}%
\end{figure*}

In this appendix we report detail results of $A_\mathrm{200c}$ fit.  To fit this, and, similarly, also $\beta$ and $\sigma$ for both $\Delta_\mathrm{200c}$ and  $\Delta_\mathrm{vir},$ as in  Eqs. \eqref{eq:A200c} and \eqref{eq:B200c}, we maximised a Likelihood as follows:

\begin{equation}
    \mathrm{ln} \mathcal{L}\left(\mathrm{ln}A_0, e_\mathbf\theta, \sigma\right) = -\frac{1}{2}\sum_i\left( \ln(2\pi\sigma^2)  + \left(\frac{\ln\ A_\mathrm{200c}\left(\mathrm{ln}A_0, e_\mathbf\theta, C_i, z_i\right)- \ln\ A_\mathrm{200c,i}}{\sigma}\right)^2\right),
\label{eq:lA200c}
\end{equation}

where $i$ runs over all setups, $C_i$ and $z_i$ are the cosmology and redshift of that setup, $A_\mathrm{200c,i}$ is the normalisation found in Section \ref{sec:ns} for a given setup (and presented in Fig. \ref{fig:hod_200c_megatabella_abs}), $e_\theta$ is a set of exponents $(e_m,e_b,e_\sigma,e_h,e_a)$  for the respective input parameters $\theta_i = (C_i,a_i)=(\Omega_{m,i},\Omega_{b,i},\sigma_{8,i},h_{0,i},a_i),$ and $\sigma$ is the fractional scatter, and $A_\mathrm{200c}$ has a power-law dependency from cosmological parameters, as follow

\begin{equation}
     \mathrm{ln}\ A_\mathrm{200c}\left(\mathrm{ln}A_0, e_\mathbf\theta, \theta_i\right) = \mathrm{ln}A_0  + \sum_i e_i\mathrm{ln}\frac{\theta_i}{\theta_{i,p}},  
\label{eq:A200cteta}
\end{equation}

where the pivot values $\theta_{i,p}$ are presented in Sec. \ref{sec:emu}.
Figure \ref{fig:posteriorA200c} shows an example of  posterior distribution of parameters $($lnA$_0,e_\theta,\sigma)$  for a single snapshot.

The evaluation of $\beta_\textrm{200c},\sigma_\textrm{200c},A_\textrm{vir},\beta_\textrm{vir}$ and $\sigma_\textrm{200c}$ have been performed in the same way as described above.

\section{Comparison with other simulation suites}
\label{ap:compa}

\begin{figure*}
   \centering
   \includegraphics[width=.9\textwidth]{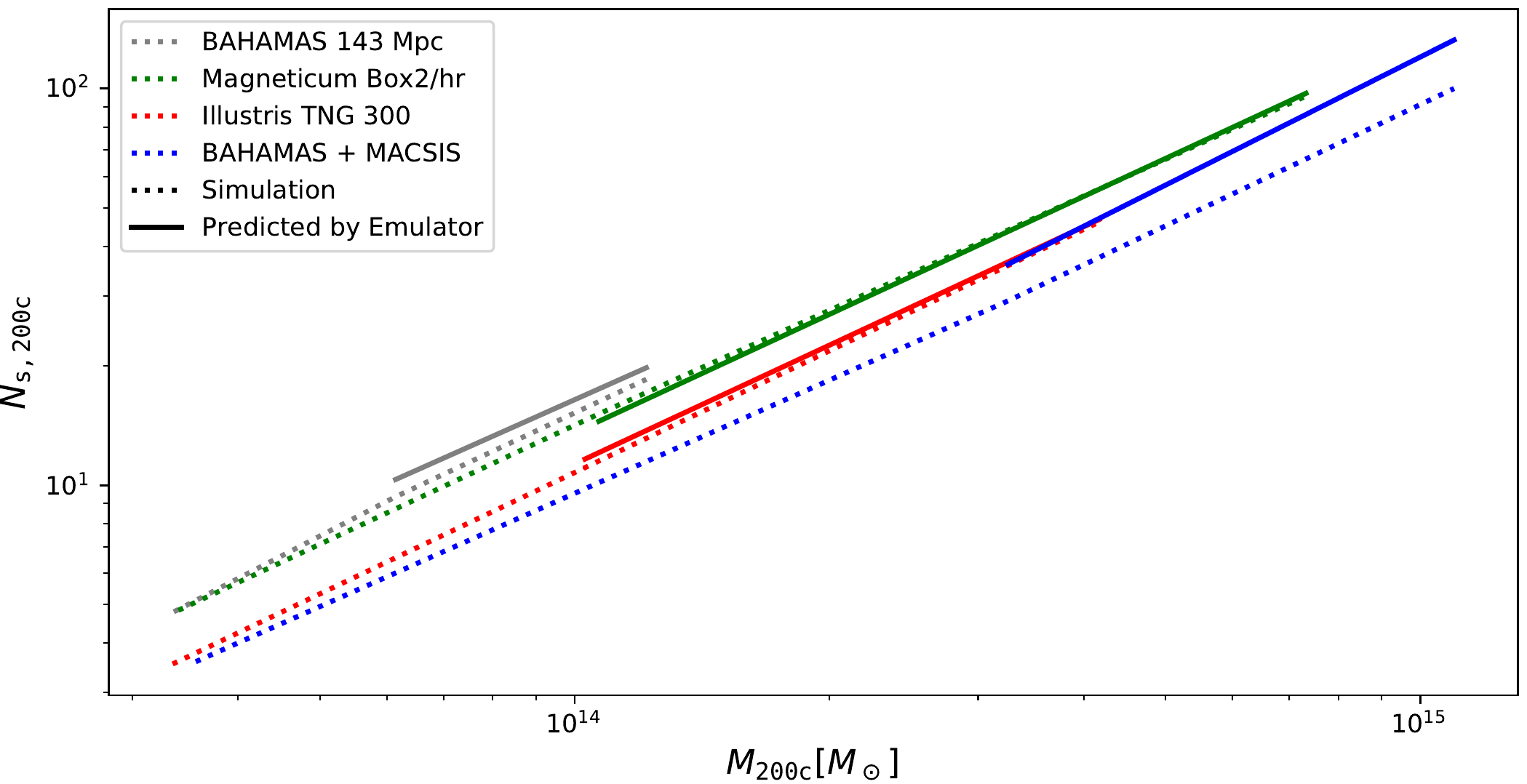}
   \caption{Dotted lines show number of satellites as a function of halo mass as taken from Fig. 2 of \protect\cite{2020MNRAS.495..686Anbajagane} for simulations BAHAMS  high resolution (BM100, gray dashed line),  \magneticum  Box2b/hr (green dashed line), IllustrisTNG 300 (red dashed line) and BAHAMAS+MACSIS (gray dashed line). Solid lines are the same cosmologies as predicted by our emulator rescaled with Eq. \eqref{eq:Nsr}. Normalisation of dashed lines is ordered as in the legend. Inset shows the distribution of Magneticum Box2/hr satellites and the respective prediction from $100$ realisations for each  halo mass, both histogram refers to a mass range $10^{14}M_\odot<M_{\rm 200c}<1.2\cdot10^{14}M_\odot$.}
    \label{fig:anbajagane_rescaled}%
\end{figure*}

In this appendix we compare the results of our emulator against the mass-richness relations of some hydrodynamic simulations in the literature.
\cite{2020MNRAS.495..686Anbajagane} (see their Table 1 for more information) compared four hydrodynamic full-physics  cosmological simulations: IllustrisTNG 300~\citep{2018MNRAS.475..648Pillepich,2018MNRAS.475..624Nelson,2018MNRAS.475..676Springel,2018MNRAS.480.5113Marinacci}, BAHAMAS 143Mpc \citep{2017MNRAS.465.2936McCarthy}, BAHAMAS+MACSIS \citep{2017MNRAS.465..213Barnes} and \magneticum Box2/hr \citep{2014MNRAS.442.2304Hirschmann,2009MNRAS.399..497Dolag} and found a difference between $\langle N_s \rangle_M$ normalisation and slopes (see their Figure 2).
In particular, their normalisation can differ of a factor $1.5,$ with BAHAMAS 143Mpc having the highest normalisation, Magneticum and BAHAMAS+MACSIS have similar slopes but different normalisation and IllustrisTNG has the highest  log-slope and the only one that  is steeper than unity.

Since these simulations have all different cosmological parameters, in this subsection we use the Emulator to test if the differences of satellite HODs of various simulations can be accounted, at least partially, by differences in their cosmological parameters.

In fact, BAHAMAS+MACSIS has parameters $\Omega_m=0.3175, f_b=0.154, \sigma_8=0.83,$ BAHAMAS has parameters $\Omega_m=0.279, f_b=0.166, \sigma_8=0.82$ and a redshift $z\approx0.1,$ IllustrisTNG 300 has parameters $\Omega_m=0.31, f_b= 0.16, \sigma_8=0.82,$ and MAGNETICUM has WMAP7 cosmological parameters.

Fig. \ref{fig:anbajagane_rescaled}  shows the $N_\textrm{s}$ $vs.$ halo mass of the four cosmological simulations and as predicted by our emulator at $z=0,$ when necessary, re-scaled for the different stellar mass cut according to Eq.~\eqref{eq:Nsr}).
We show the emulator prediction only for the high-mass power law regime.

{The emulator  {matches} the normalisation of IllustrisTNG and BAHAMS 143 Mpc.}
On the other hand, the emulator over-predicts BAHAMAS+MACSIS normalisation.

\end{appendix}

\end{document}